\documentclass[aps,prl,twocolumn]{revtex4}
\pdfoutput=1
\usepackage{graphicx}
\usepackage{amsmath}
\usepackage{amssymb}
\usepackage{color}
\usepackage{bm}

\definecolor{red}{rgb}{0.75,0,0}
\definecolor{blue}{rgb}{0,0,0.75}
\definecolor{green}{rgb}{0,0.5,0}

\newcommand{\revision}[1]{#1}

\DeclareMathOperator{\asinh}{asinh}
\DeclareMathOperator{\acosh}{acosh}
\DeclareMathOperator{\atan}{atan}
\DeclareMathOperator{\sech}{sech}

\begin{document}
\title{Hyperbolic Interfaces}

\author{Luca Giomi}
\affiliation{School of Engineering and Applied Sciences, Harvard University, Cambridge, MA 02138,  USA}

\date{\today}

\begin{abstract}
Fluid interfaces, such as soap films, liquid droplets or lipid membranes, are known to give rise to several special geometries, whose complexity and beauty continue to fascinate us, as observers of the natural world, and challenge us as scientists. Here I show that a special class of surfaces of constant negative Gaussian curvature can be obtained in fluid interfaces equipped with an orientational ordered phase. These arise in various soft and biological materials, such as nematic liquid crystals, cytoskeletal assemblies, or hexatic colloidal suspensions. The purely hyperbolic morphology originates from the competition between surface tension, that reduces the area of the interface at the expense of increasing its Gaussian curvature, and the orientational elasticity of the ordered phase, that in turn suffers for the distortion induced by the underlying curvature.
\end{abstract}

\maketitle

Understanding the origin of shape in structures at various length scales is a central goal across many areas of science and engineering. Often, the shape is dictated by the balance of competing forces simultaneously acting on a system, or by the interplay between spontaneous geometry and the residual stresses accumulated during fabrication and growth \cite{Liang:2011}.  A standard problem consists of finding the optimal shape of a system given its mechanical energy and the constraints the system is subject to. Conversely, given a target shape with specific geometrical properties, one could ask what is the simplest physical system in which this will occur as a result of mechanical equilibrium and kinetics. Such an {\em inverse problem} represents a crucial step in several aspects of modern technology, ranging from robotics \cite{Kofod:2007} to medicine \cite{Euliss:2006}. In the case of soft elastic solids, such as thin sheets of polymer gels, controlling the amount of local growth or swelling was demonstrated to be a powerful technique to produce various three-dimensional shapes of both positive and negative Gaussian curvature \cite{Klein:2007}. Fluid interfaces also bear a great potential for designing surfaces with special or tunable geometric properties. Soap films, droplets, vesicles, micelles and membranes are examples of this potential as well as of the intrinsic beauty of these fragile objects. 

When endowed with in-plane orientational order, the richness of shapes that can be attained by fluid interfaces becomes enormous. In a recent article Xing {\em et al}. \cite{Xing:2012} have shown that nematic and smectic vesicles formed from block copolymers with liquid-crystalline side chains, self-assemble in a great variety of morphologies due to the competition between in-plane liquid-crystalline order and bending elasticity. In a spectacular experimental work, Gibaud {\em et al}. \cite{Gibaud:2012} have reported the formation of several complex structures in colloidal membranes, arising from the interplay between chiral order and interfacial tension. The morphogenetic force in these systems originates from the fact that orientational order is {\em frustrated} by the Gaussian curvature of the underlying surface \cite{Bowick:2009,Turner:2010,Mbanga:2012}. When parallel transported on a surface of non-zero Gaussian curvature, a tangent vector rotates with respect to its original orientation and the amount of rotation is proportional to the Gaussian curvature experienced along the path. For this reason, local orientational order, whether nematic, smectic or chiral, is inevitably hindered in presence of Gaussian curvature.

In this Letter I show that a special class of surfaces of constant negative Gaussian curvature, can be obtained in fluid interfaces equipped with an orientational ordered phase. This remarkable phenomenon originates from the competition between surface tension, that reduces the area of the interface at the expense of increasing its Gaussian curvature, and the orientational elasticity of the ordered phase, that in turn suffers for the distortion induced by the underlying Gaussian curvature. Hints of this behavior were observed by Frank and Kardar \cite{Frank:2008}, who reported an example of defect-induced buckling of a nematic membrane into the pseudosphere. The latter can be considered a particular manifestation of a more general phenomenon that, as we will see in the following, does not require defects nor any specific rotational symmetry of the ordered phase.     
 
Let us consider a two-dimensional interface $M$ of surface tension $\sigma$, equipped with an orientationally ordered phase. This can arise in various soft materials, such as nematic liquid crystals, tilted molecular chains, or hexatic suspensions. Orientational order on a surface can be described via a unit vector $\bm{n}$ in the tangent plane of the surface. The total energy of the system can then be expressed in the form:
\begin{equation}\label{eq:energy}
E = \int dA\,(\sigma + \kappa\,|\nabla\bm{n}|^{2})\;,	
\end{equation}  
where $\nabla$ indicates the covariant derivative in the metric of $M$, and $\kappa$ is an orientational stiffness constant with dimensions of energy. The orientational contribution in Eq. \eqref{eq:energy} describes, the one elastic constant approximation, a generic $p-$atic phase, where local orientations are defined modulo $2\pi/p$ \cite{Bowick:2009}. $p=1$ represents polar order, $p=2$ corresponds to a nematic phase and $p=6$ characterizes hexatic interfaces. The energy has a characteristic length scale, in addition to that set by the boundary, given by $\ell=\sqrt{\kappa/\sigma}$. Thus we might expect the bulk Gaussian curvature of the interface to be of order $K=-\ell^{-2}$, with the negative sign due to the natural hyperbolicity of fluid interfaces. Remarkably, this simple estimate is {\em exact} within a range of material and geometrical parameters, resulting in an interface of constant negative Gaussian curvature, whose intrinsic geometry is that of the hyperbolic plane. The remainder of this paper will be devoted to prove this result and explore its consequences in the special case of tubular interfaces.

As a starting point, we express the vector field $\bm{n}$ in a local orthonormal frame of tangent vectors $\bm{e}_{\alpha}$ ($\alpha=1,\,2$), so that: $\bm{n}=\cos\theta\,\bm{e}_{1}+\sin\theta\,\bm{e}_{2}$. Using standard manipulations \cite{David:1989,Bowick:2009}, the gradient terms in Eq. \eqref{eq:energy} can be then expressed as $\nabla_{i}\bm{n}=(\partial_{i}\theta-A_{i})\,\bm{n}_{\perp}$, where $A_{i}=\bm{e}_{1}\cdot\partial_{i}\bm{e}_{2}$ is the {\em spin connection} that accounts for the rotation of the local frame $\bm{e}_{\alpha}$ as we move along the surface and $\bm{n}_{\perp}=\cos\theta\,\bm{e}_{2}-\sin\theta\,\bm{e}_{1}$. The Gaussian curvature of the surface is given by the curl of the spin connection: $K=\epsilon^{ij}\nabla_{i}A_{j}$, with $\epsilon^{ij}$ the antisymmetric Levi-Civita tensor. As a consequence, the orientational contribution to the energy \eqref{eq:energy} cannot vanish everywhere because $\bm{A}$ cannot equate the curl-free vector field $\nabla\theta$ on a surface having non-zero Gaussian curvature \cite{Turner:2010}. Orientational order is therefore geometrically frustrated.

Minimization of Eq. \eqref{eq:energy} with respect to the orientation field $\theta$, leads to the well known equation \cite{Vitelli:2004}:
\begin{equation}\label{eq:angle}
\nabla\cdot(\nabla\theta-\bm{A}) = 0\;,
\end{equation}
subject to the constraint:
\begin{equation}\label{eq:constraint}
\epsilon^{ij}\nabla_{i}(\partial_{j}\theta-A_{j}) = \eta-K\;,
\end{equation}
where $\eta$ is the topological charge density, should topological defects be present in the configuration of $\bm{n}$ \cite{Bowick:2009}. The orientation at the boundary $\partial M$ can either be prescribed, in which case $\theta=\theta(s)$ at $\partial M$ (with $s$ the arc-length of the boundary curve), or free. In the latter case energy minimization further demands $\bm{\nu}\cdot(\nabla\theta-\bm{A})=0$ at the boundary. Here $\bm{\nu}$ is the outward directed tangent vector normal to the boundary curve $\partial M$. A classic strategy to solve Eq. \eqref{eq:angle} under the constraint \eqref{eq:constraint}, is to express the vector field $\nabla\theta-\bm{A}$ through the antisymmetrized derivatives of a scalar field $\varphi$: $\partial_{i}\theta-A_{i}=\epsilon_{i}^{k}\partial_{k}\varphi$. With this choice, Eq. \eqref{eq:angle} is automatically satisfied, while Eq. \eqref{eq:constraint} implies:
\begin{equation}\label{eq:poisson}
\Delta_{g}\varphi = \eta-K \;,	
\end{equation}
where $\Delta_{g}=\nabla_{i}\nabla^{i}$ is the Laplace-Beltrami operator in the metric of $M$. As it was observed by Vitelli and Nelson \cite{Vitelli:2004}, the scalar field $\varphi$ plays the role of a {\em geometric potential} expressing the amount of non-uniformity in the vector field $\bm{n}$ caused by the presence of topological charge, whether localized in the defects or distributed in the form of Gaussian curvature. In terms of $\varphi$, the energy \eqref{eq:energy} of the interface can be rewritten in the simple form:
\begin{equation}\label{eq:energy-phi}
E = \int dA\,(\sigma+\kappa\,|\nabla\varphi|^{2})\;.	
\end{equation}
In the case of free orientation at the boundary, $\bm{t}\cdot\nabla\varphi=-\bm{\nu}\cdot(\nabla\theta-\bm{A})=0$, hence $\varphi$ is constant along the boundary curve and can, in particular, be set to zero as the energy \eqref{eq:energy-phi} only depends on the gradient of $\varphi$. Analogously, if $\bm{n}$ is everywhere {\em tangent} to $\partial M$, $\bm{\nu}\cdot\nabla\varphi=\bm{t}\cdot(\nabla\theta-\bm{A})=\kappa_{g}$, where $\kappa_{g}$ is the geodesic curvature of the boundary \cite{Kamien:2002}. With the help of the Gauss-Bonnet theorem \cite{Kamien:2002,Gray:1997}, however, one can prove that this latter condition is automatically satisfied by any solution of Eq. \eqref{eq:energy-phi}, thus we can again take $\varphi=0$ at $\partial M$.

Next, we calculate the energy variation corresponding to a displacement of the interface along the normal direction: $\bm{r}\rightarrow\bm{r}+\epsilon\bm{N}$, with $\bm{R}$ the position vector and $\epsilon$ a small displacement along the normal vector $\bm{N}$. The variation of the area element is a classic result of differential geometry: $\delta(dA)=-2H\epsilon\,dA$ with $H$ the surface mean curvature \cite{Gray:1997}. In order to calculate the variation of the orientation energy, we make the assumption that the system is defect free, so that $\eta=0$ and the second term in \eqref{eq:energy-phi} can be expressed as:
\[
W = \int dA\,|\nabla\varphi|^{2}= - \int dA\,dA'\,G(\bm{r},\bm{r}')K(\bm{r})K(\bm{r}')\;, 	
\]
where $G(\bm{r},\bm{r}')$ is the Laplacian Green function, with $G(\bm{r},\,\cdot\,)=G(\,\cdot\,,\bm{r})=0$ for $\bm{r}\in\partial M$. Next, we introduce a system of {\em conformal coordinates} $(u^{1},u^{2})$, so that the metric is locally Euclidean up to a space dependent scaling factor: $ds^{2}=w[(du^{1})^{2}+(du^{2})^{2}]$ \cite{Bowick:2009}. In conformal coordinates, the area element is given by $dA=w\,du^{1}du^{2}$ and the Gaussian curvature can be expressed in the simple form:
\begin{equation}\label{eq:gaussian}
K = -\frac{\Delta \log w}{2w}\;,
\end{equation}
where $\Delta=w\Delta_{g}=\partial_{u_{1}}^{2}+\partial_{u_{2}}^{2}$ is the Euclidean Laplacian. In the new coordinates the functional $W$ becomes:
\[
W = -\frac{1}{4}\int d^{2}u\,\Delta \log w(\bm{u}) \int d^{2}u'\,G(\bm{u},\bm{u}')\,\Delta \log w(\bm{u}') \;, 
\]
where $d^{2}u=du^{1}du^{2}$. With these simplifications, all the geometrical structure of the interface is embodied in the conformal weight $w$ and the expression of $W$ is now suitable to calculate the normal variation. Consistently with the area variation given above, one has that $\delta w = -2wH\epsilon$ and $\delta(\log w)=-2H\epsilon$. Thus, after some manipulations, we find:
\[
\delta W = \int d^{2}u\,\Delta(H\epsilon) \int d^2u'\,G(\bm{u},\bm{u}')\,\Delta\log w(\bm{u}')\;. 
\] 
Now, the integral over $\bm{u}'$ is a function of $\bm{u}$ that vanishes when $\bm{u}\in\partial M$. Thus, using Green's identities and taking into account that $\epsilon=0$ at the boundary, we find:
\[
\delta W = \int d^{2}u\, H\epsilon\,\Delta\log w(\bm{u}) = -2\int dA\,KH\epsilon\;.
\] 
Combining this with the area variation we finally obtain an expression for the normal variation of the total energy:
\begin{equation}\label{eq:total-variation}
\delta E = \int dA\,(\sigma+\kappa K)(-2H\epsilon)\;. 	
\end{equation}
Evidently, the variational equation $\delta E=0$ has two solutions: $H=0$ and, as anticipated, $K=-\sigma/\kappa$. This implies that, in order for the interface to have minimal energy, it can either form a surface of zero mean curvature, such as a classic soap film, or a surface of constant negative Gaussian curvature. \revision{In presence of topological defects, the Gaussian curvature in Eq. \eqref{eq:total-variation} is replaced by $K-\eta$ and energy is again minimized by setting $K=-\sigma/\kappa$ outside of the defect-core \cite{Frank:2008}.} As we observed in the introduction, this behavior originates from the competition between surface tension, that acts to reduce the area of the interface with a consequent increase in curvature, and the orientational elasticity of the order phase, that in turn is subject to a stronger distortion as the Gaussian curvature is increased. For this reason, we might also expect that the stability of the purely hyperbolic shape would be lost for small values of $\kappa$, when the distortion induced by the curvature of the interface is energetically inexpensive. In the remainder of the paper, we will consider the example of a tubular interface to clarify these assertions.

\begin{figure}
\centering
\includegraphics[width=0.6\columnwidth]{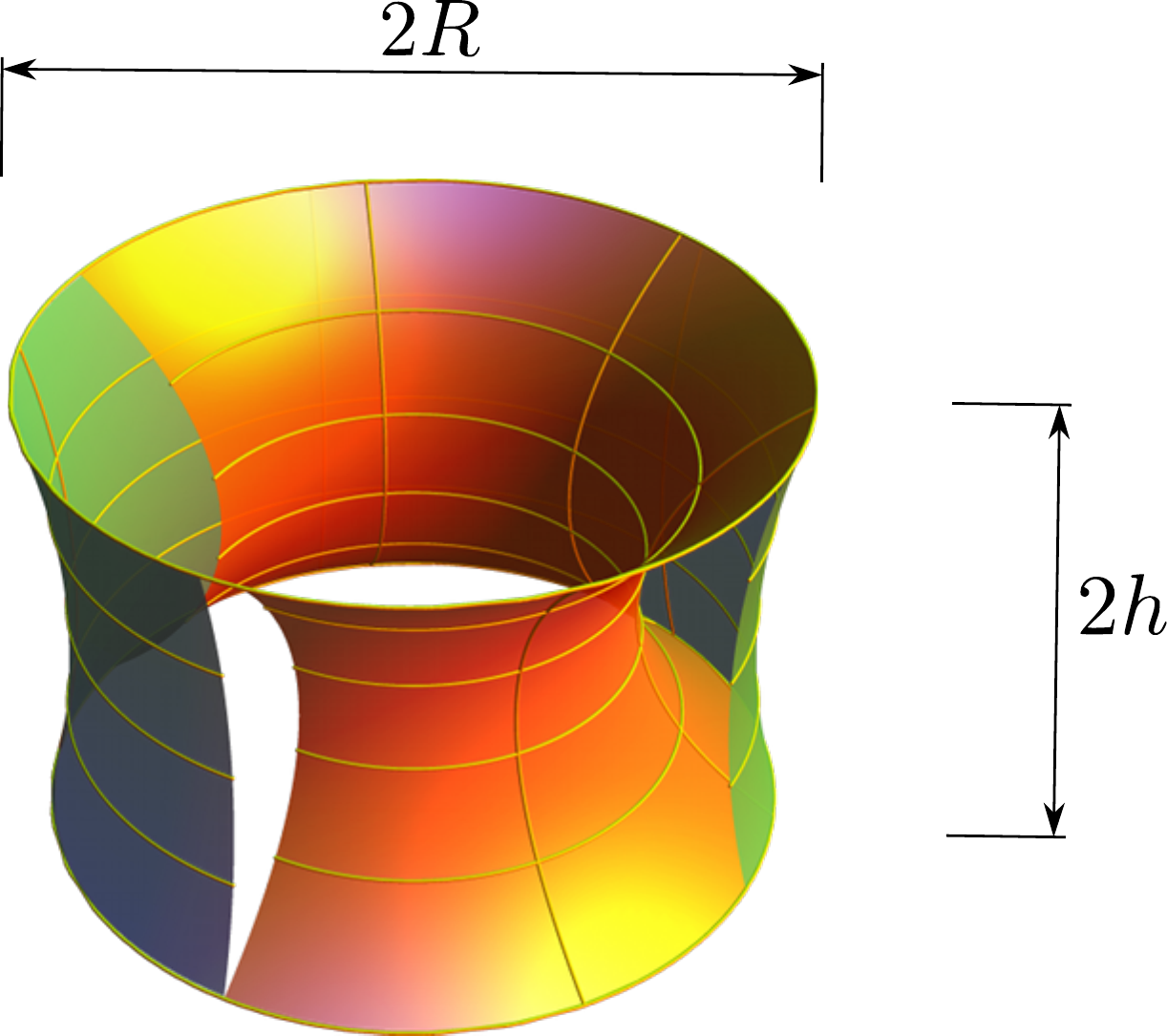}
\caption{\label{fig:surface}A portion of catenoid (orange) superimposed by a portion of the pseudospherical hyperboloid described in text (green). Both surface are bounded by a pair of rims of radius $R$ held at a distance $2h$ from each other.}	
\end{figure}

Let us consider an interface bounded by two circular rims of radius $R$ held at distance $2h$ from each other (Fig. \ref{fig:surface}). In absence of any orientational order, the shape of the interface is given by the classic catenoid, parametrized as:
\[
x = a \cosh \left(\frac{u}{a}\right) \cos\phi\;,\quad
y = a \cosh \left(\frac{u}{a}\right) \sin\phi\;,\quad
z = u\;,
\]
where $0\le \phi < 2\pi$ is the usual polar angle and $-h\le u \le h$. The mean curvature is everywhere zero while the Gaussian curvature is given by $K=-\sech^{4}\left(u/a\right)/a^{2}$. The length $a$ is obtained by setting the polar radius to $R$ when $z=\pm h$. This yields $R=a\cosh(h/a)$. It is well known that the solution of this transcendental equation exists only if $h\le R/\min_{x}\{\cosh(x)/x\}\simeq 0.663\,R$, while for $h/R>0.663$ the catenoid does not exist. In the $(u,\phi)$ coordinates the spin connection has components: $A_{u}=0$ and $A_{\phi}=-\tanh(u/a)$.

For large orientational stiffness we expect the interface bounded by the two circular rims to be a surface of revolution of constant negative Gaussian curvature. A surface with this properties can be identified in the pseudospherical hyperboloid \cite{Gray:1997}, whose parametric form is given by:
\begin{gather*}
x = b \cosh \left(\frac{u}{a}\right) \cos\phi\;,\quad
y = b \cosh \left(\frac{u}{a}\right) \sin\phi\;,\\[5pt]
z = -iaE\left(\frac{iu}{a}\bigg|-\frac{b^{2}}{a^{2}}\right)\;,
\end{gather*}
with $0\le \phi < 2\pi$ and $-U\le u \le U$. Here $E(\phi|m)=\int_{0}^{\phi}dt\,(1-m\sin^{2}t)^{1/2}$, with $|m|\le 1$, is the incomplete elliptic integral of second kind. The coordinate $u$ is limited in magnitude $|u| \le a\asinh(a/b)$ and the surface has constant negative Gaussian curvature $K=-1/a^{2}$. This immediately fixes the value of the constant $a=(\kappa/\sigma)^{1/2}$. The length $b$, on the other hand, is found by requesting $z=\pm h$, when $u=\pm U=\pm a\acosh(R/b)$. This gives:
\begin{equation}\label{eq:h}
h = -ia E\left(i\acosh \frac{R}{b} \bigg| -\frac{b^{2}}{a^{2}}\right)\;.
\end{equation}
The solution of this equation must satisfy $a \ge b$, $R \ge b$ and $R^{2}\le a^{2}+b^{2}$ as the consequence of the bounds on the coordinate $u$ and the elliptic modulus. The components of the spin connection are: $A_{u}=0$ and $A_{\phi}=-(b/a)\sinh(u/a)$.

\begin{figure}[t]
\centering
\includegraphics[width=0.7\columnwidth]{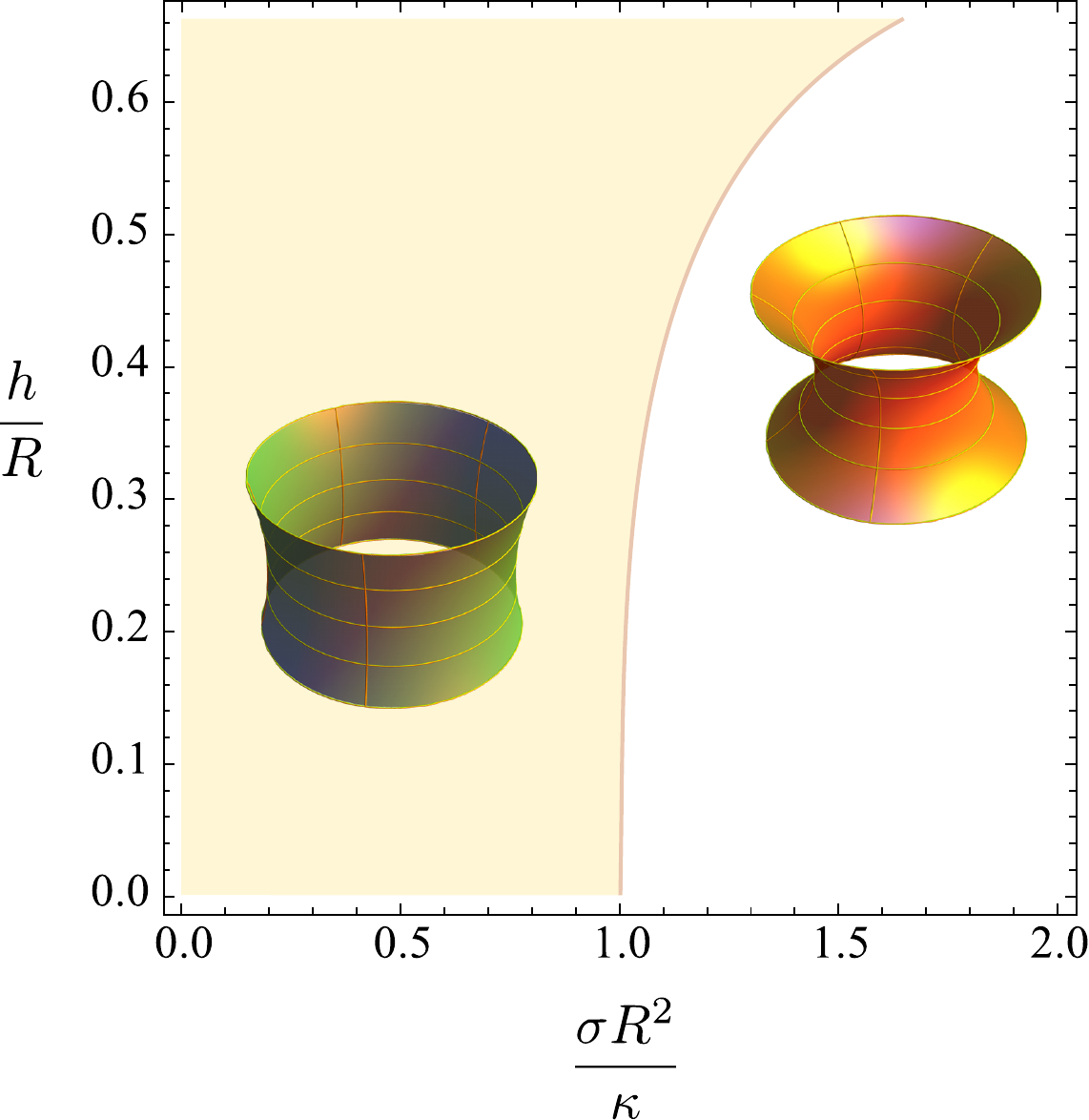}
\caption{\label{fig:phase-diagram}Phase-diagram obtained by comparing the energies in Eqs. \eqref{eq:energy-surfaces}. For $\sigma R^{2}/\kappa>1$ the equilibrium shape is a catenoid. $\sigma R^{2}/\kappa<1$, on the other hand, the energy balance is dominated by the orientational stiffness of the ordered phase and the interface is a surface of constant negative Gaussian curvature $K=-\sigma/\kappa$.}	
\end{figure}

Now, because both the catenoid and the pseudospherical hyperboloid are azimuthally symmetric, all the relevant quantities will be independent on $\phi$, in particular $\theta=\theta(u)$ and $\varphi=\varphi(u)$. This latter property implies that $\theta$ is constant throughout the interface since: $\partial_{u}\theta = A_{u}+\epsilon_{u}^{\phi}\partial_{\phi}\varphi = 0$. Thus the orientational energy density simplifies to $|\bm{A}|^{2}=g^{\phi\phi}A_{\phi}^{2}$ ($g_{ij}$ being the metric tensor). The total energies can be readily calculated in the form:
\begin{subequations}\label{eq:energy-surfaces}
\begin{gather}
\frac{E_{\rm cat}}{2\pi\kappa} = 2\left(\frac{h}{a}-\tanh \frac{h}{a}\right)+\frac{\sigma a}{\kappa}\left(h+\frac{a}{2}\sinh\frac{2h}{a}\right)\;, \\[5pt]
\frac{E_{\rm hyp}}{2\pi\kappa} = \frac{2b}{a}\left(\chi-\atan\chi\right)+\frac{2\sigma ab\,\chi}{\kappa}\;,
\end{gather}
\end{subequations}
where $\chi=\sqrt{(R/b)^{2}-1}$. As a consistency check, one could also verify that the geometric potential $\varphi$ is given for {\em both} surfaces by the simple function: $\varphi = \log(r/R)$, with $r$ the usual polar distance in the $xy-$plane. Substituting in Eq. \eqref{eq:energy-phi} and integrating, promptly gives Eqs. \eqref{eq:energy-surfaces}. A comparison of the energies \eqref{eq:energy-surfaces} leads to the diagram of Fig. \ref{fig:phase-diagram} in the plane $(\sigma R^{2}/\kappa,h/R)$. As expected for $\sigma R^{2}/\kappa>1$, surface tension dominates and the interface has the classic shape of a catenoid. For $\sigma R^{2}/\kappa<1$, on the other hand, the energy balance is dominated by the orientational stiffness of the ordered phase and the interface adapts to the in-plane orientational order by forming a surface of constant negative curvature. The stability of this purely hyperbolic interface is enhanced when the aspect ratio $h/R$ is closer to the upper bound $h/R\sim 0.663$. 

A check of the orders magnitude is inevitable at this point. Standard liquid crystals have $\kappa\sim k_{B}T\sim 10^{-14}$ erg at room temperature, whereas the typical surface tension of liquid crystal interface is of order $\sigma\sim 10$ erg/cm$^2$. Thus, having $\sigma R^{2}/\kappa\sim 1$ would require a rim of radius $R<1$ nm, which is smaller than a typical film thickness. Block copolymer membranes, as those considered in Ref. \cite{Xing:2012}, are more suitable candidates to observe the hyperbolic morphology described here, as the orientational stiffness is of order $60$ $k_{B}T$. In actomyosin compounds, the contractile forces generated by the molecular motors have been recently demonstrated to give rise to an {\em effective} surface tension \cite{Mertz:2012}. Due to their elongated shape as well as the cross-linking operated by the myosin motors and, possibly, by other actin-binding-proteins, actin filaments can form a nematic phase at room temperature. Since the forces exerted by the motors, of order of pico-Newtons, are in this case responsible for both the surface tension and the alignment, one can expect their effect to be comparable in magnitude so that the balance $\sigma R^{2}/\kappa\sim 1$ could be achieved for reasonable system sizes.

\revision{Hexatic suspensions obtained by trapping colloidal particles at the interface between two liquids, represent an especially promising candidate for an experimental realization the phenomenon discussed in this Letter. In this case the orientational stiffness can be finely controlled by moving across the liquid/hexatic/solid phase diagram. As predicted in the theory of two-dimensional melting \cite{Nelson:2002} the orientational stiffness has a universal value $\kappa/k_{B}T=72/\pi$ at the liquid/hexatic transition and {\em diverges} exponentially at the hexatic/solid transition. This was brilliantly verified by Keim {\em et al}. \cite{Keim:2007} in experiments with superparamagnetic colloidal particles with tunable interaction strength at the water/air interface.  Thus, by approaching the hexatic/solid transition, one could in principle make the orientational stiffness arbitrarily large ($\sim 10^{3}$ $k_{B}T$ in practice). In turn, the interfacial tension between two immiscible liquid phases can be made arbitrarily small as the critical point for phase separation is approached.}

In conclusion, I showed that surfaces of constant negative Gaussian curvature can be realized in fluid interfaces equipped by an orientational ordered phase as a consequence of the interplay between surface tension and orientational elasticity. A simple example of this behavior was presented in the case of defect-free tubular interfaces. \revision{An experimental realization of this phenomenon could be achieved in hexatic colloidal suspensions and remains a challenge for the future.}

\acknowledgments
This work is partially supported by the NSF Harvard MRSEC, the Harvard Kavli Institute for Nanobio Science \& Technology and the Wyss Institute. I am grateful to Alessandra Silvestri for useful suggestions.

\end{document}